\documentclass[12pt,a4paper]{article}
\usepackage{verbatim} 
\usepackage{jheppub} 

\usepackage[T1]{fontenc} 
\usepackage[utf8]{inputenc}
\usepackage{amsfonts,amssymb,amsmath,amsthm}
\usepackage{graphicx}
\usepackage{slashed}

\newcommand{\MM}{\mathcal{M}}
\newcommand{\ii}{\mathrm{i}}
\newcommand{\nn}{\mathbf{n}}

\newcommand{\dd}{\mathrm{d}}

\newcommand{\be}{\begin{equation}}
\newcommand{\ee}{\end{equation}}
\newcommand{\bea}{\begin{eqnarray}}
\newcommand{\eea}{\end{eqnarray}}

\newtheorem{assumption}{Assumption}[section]

\newtheorem{definition}[assumption]{Definition}

\title{Anomaly inflow for local boundary conditions}

\author[a,b]{A. V. Ivanov}
\emailAdd{regul1@mail.ru}
\author[c]{and D. V. Vassilevich}
\emailAdd{dvassil@gmail.com}

\affiliation[a]{St. Petersburg Department of Steklov Mathematical Institute of
RAS,\\ 27 Fontanka, St. Petersburg 191023, Russia}
\affiliation[b]{Leonhard Euler International Mathematical Institute,\\ 10 Pesochnaya nab., St. Petersburg 197022, Russia}
\affiliation[c]{CMCC-Universidade Federal do ABC,\\ Avenida dos Estados 5001, CEP 09210-580, Santo Andr\'e, S.P. Brazil}

\abstract{We study the $\eta$-invariant of a Dirac operator on a manifold with boundary subject to local boundary conditions with the help of heat kernel methods. In even dimensions, we relate this invariant to $\eta$-invariants of a boundary Dirac operator, while in odd dimension, it is expressed through the index of boundary operators. We stress the necessity of the strong ellipticity condition for the applicability of our methods. We show that the Witten--Yonekura boundary conditions are not strongly elliptic, though they are very close to strongly elliptic ones.}

\begin{document}

\maketitle

\section{Introduction}\label{sec:in}
The idea of an anomaly inflow, which roughly means that some anomalies can be related to the anomalies in effective theories on boundaries or domain walls, can be traced back to the paper \cite{Callan:1984sa}. In this paper, as well as in the present work, the following two anomalies were considered. The first one was a global version of the Adler--Bell--Jackiw  \cite{Adler:1969gk,Bell:1969ts} anomaly, which is related to the index of a Dirac operator. The second one was the parity anomaly 
\cite{Redlich:1983dv,Niemi:1983rq,AlvarezGaume:1984nf} defined by the $\eta$-invariant. An extremely important related development in mathematics appeared even earlier, in the paper by Atiyah, Patodi, and Singer (APS) \cite{Atiyah:1975jf}, where a relation between the index of a Dirac operator and the $\eta$-invariant of an auxiliary Dirac operator on the boundary was established. Applications of this result to topological phases of matter for a generalization of APS problem were studied in \cite{Yonekura:2016wuc} basing on a previous work by Dai and Freed \cite{Dai:1994kq}. Note that the APS boundary conditions are non-local, which is not very convenient for physical applications.

Recently, theorems analogous to the APS Index Theorems were established for domain wall configurations. In \cite{Vassilevich:2018aqu,Ivanov:2020fsz,Ivanov:2020rug} the domain walls were the surfaces, where the gauge connection is discontinuous, while in \cite{Fukaya:2019qlf,Fukaya:2020tjk,Onogi:2021slv} they were related to discontinuities of the mass function.

Let us turn to local boundary conditions. The parity anomaly for Euclidean bag boundary conditions in 4 dimensions was computed in \cite{Kurkov:2017cdz,Kurkov:2018pjw} by using the heat kernel methods. It was demonstrated that for a vanishing extrinsic curvature of the boundary, this anomaly is equal to one half of the parity anomaly for a boundary Dirac operator. Extensions of these results to chiral bag boundary conditions can be found in \cite{Ivanov:2021yms}. The problem with Euclidean (chiral) bag conditions is that they can be formulated on even-dimensional manifolds only. The anomaly inflow for local boundary conditions existing in any number of dimensions was studied by Witten and Yonekura (WY) \cite{Witten:2019bou}. Their boundary conditions lead to a non-hermitian Dirac operator. Nevertheless, the authors managed to establish relations between the phase of the Dirac determinant and the $\eta$ function of an operator on the boundary by using a kind of Pauli--Villars regularization procedure. It is interesting to check whether the heat kernel methods may be applied to these conditions as well. It is also interesting to find a set of local boundary conditions in odd dimensions, which lead to a hermitian Dirac operator and admit an anomaly inflow. In this paper, we address both problems mentioned above and extend the results for bag boundary conditions obtained in 4D to arbitrary even-dimensional manifolds.

To complete a small panorama of the field, we would like to mention a work  \cite{Hsieh:2020jpj}, which extends anomaly inflows to $p$-form gauge theories, and a very recent preprint \cite{Davighi:2022icj}, which studies anomalies of non-abelian finite groups in this formalism. 

Let $\MM$ be a compact smooth oriented spin manifold with a smooth boundary $\partial\MM$ and let $V$ be a spin and gauge bundle over $\MM$. The sections of $V$ are spinor fields denoted by $\psi$ in what follows. The Dirac operator
\begin{equation}
\slashed{D}=\ii \gamma^\mu \nabla_\mu\label{Dirop}
\end{equation}
acts on sufficiently smooth spinor fields. The $\gamma$-matrices satisfy $\gamma^\mu\gamma^\nu +\gamma^\nu\gamma^\mu=2g^{\mu\nu}$. In a local basis, the covariant derivative can be written as
\begin{equation}
\nabla_\mu = \partial_\mu + \tfrac 18 w_{\mu\nu\rho}[\gamma^\nu,\gamma^\rho ]+A_\mu, \label{nabla}
\end{equation}
where $w_{\mu\nu\rho}$ is a spin connection, while $A_\mu$ is a gauge field.

Let $\lambda$ denotes the eigenvalue of $\slashed{D}$. Then the $\eta$ function of $\slashed{D}$ is defined by an infinite sum
\begin{equation}
\eta(s,\slashed{D})=\sum_{\lambda>0}\lambda^{-s}-\sum_{\lambda<0} (-\lambda)^{-s},
\label{eta}
\end{equation}
where $s$ is a complex parameter.
If the operator $\slashed{D}$ has complex eigenvalues, the conditions $\lambda>0$ and $\lambda<0$ in (\ref{eta}) are replaced by the conditions $\Re\lambda>0$ and $\Re\lambda<0$, respectively.

In a "normal" situation, the sum on the right hand side of Eq.\ (\ref{eta}) is convergent, when $\Re s$ is sufficiently large, and defines a meromorphic function of $s$ on the whole complex plane. Then, one can say that the value of $\eta(0,\slashed{D})$ measures an asymmetry of the spectrum of $\slashed{D}$. If the spectrum is real, $\eta(0,\slashed{D})$ defines the phase of $\det(\slashed{D})$ and the parity anomaly. If there are complex eigenvalues, physical interpretation of $\eta(0,\slashed{D})$ is less clear.

"Normality" mentioned in the previous paragraph depends on two requirements: the operator has to be elliptic, and the boundary value problem has to be strongly elliptic. Roughly speaking, the ellipticity means that the leading symbol (which is just a Fourier transform of the part of an operator containing the highest order derivatives) has to be non-degenerate for a non-vanishing wave vector $k_\mu$. For the Dirac operator, the leading symbol is $p=\gamma^\mu k_\mu$. Obviously, the Dirac operator is elliptic. 

The strong ellipticity requirement is more subtle. It treats certain properties of an auxiliary boundary value problem at all points of the boundary, where one keeps the higher normal derivatives in the operator, while the tangential derivatives are replaced by wave vectors. A precise definition of the strong ellipticity will be given in Section \ref{sec:def}. The main idea of the ellipticity conditions is to ensure that the eigenvalues grow sufficiently fast and that they do not appear too far away from the real axis.

We should also mention that even with strongly elliptic boundary conditions $\eta(s,\slashed{D})$ may in principle have a pole at $s=0$.

Consider a smooth family of Dirac operators $\slashed{D}_u$ depending on a parameter $u$. When an eigenvalue $\lambda$ crosses the line $\Re \lambda=0$, the value of $\eta(0,\slashed{D}_u)$ jumps by 2. However, as long as no eigenvalue changes the sign of its real part, $\eta(0,\slashed{D}_u)$ varies smoothly. Thus, $\eta(0,\slashed{D}_u)$ is smooth mod 2.

In this work, we will consider exclusively local boundary conditions\footnote{An example of nonlocal boundary conditions for the Dirac operator is given by the Atiyah--Patodi--Singer spectral conditions \cite{Atiyah:1975jf}.} for the Dirac operator. Their general form is
\begin{equation}
\Pi_-\psi\vert_{\partial\MM}=0,\label{bc}
\end{equation}
where $\Pi_-$ is an ultralocal projector of the rank $\tfrac 12 \, \mathrm{rank}(V)$. (Ultralocality means the $\Pi_-$ is a matrix at each point of the boundary.) If boundary conditions belong to this class and they are strongly elliptic, there is a full asymptotic expansion of the heat trace
\begin{equation}
\mathrm{Tr}\, \left( Q e^{-t\slashed{D}^2}\right)\simeq \sum_{k=0}^{+\infty} t^{\frac {k-n}2}a_k(Q,\slashed{D}^2)\label{hkexp}
\end{equation}
for $t\to +0$. Here $Q$ is any smooth endomorphism of $V$, $n=\mathrm{dim}\, \MM$. The heat kernel coefficients are local. This means that they are given by sums of volume and surface integrals of local geometric invariants defined in the bulk and on the boundary, respectively. An introduction to the heat kernel techniques written for physicists can be found in \cite{Vassilevich:2003xt}.

The variations of $\eta(0,\slashed{D})$ can be expressed through the heat kernel \cite{Atiyah:1980jh,Gilkey:1984,AlvarezGaume:1984nf}. In particular, if the variation $\delta\slashed{D}$ is a zero-order operator (as, e.g., in the case, when this variation is caused by a variation of $A$), then we have
\begin{equation}
\delta\eta(0,\slashed{D})=-\frac 2{\sqrt{\pi}}\, a_{n-1}(\delta\slashed{D},\slashed{D}^2). \label{vareta}
\end{equation}
The metric variations lead to $\delta\slashed{D}$ being a first-order differential operator. How to deal with such variations is explained in \cite{Kurkov:2018pjw}, see also \cite{Branson:1997ze}. In what follows, we shall only need the fact that the corresponding variation is given again by a variation of the heat kernel coefficient, this time of $a_{n+1}$.

Locality of the heat kernel coefficients plays an important role here. It allows us to disentangle the bulk and the boundary parts in the variation of $\eta(0,\slashed{D})$. Since odd-numbered heat kernel coefficients $a_{2k+1}$ do not have bulk parts, the variations of $\eta(0,\slashed{D})$ on even dimensions have boundary contributions only. Thus, to check the anomaly inflow conjecture on even-dimensional manifolds, one should relate the boundary part of $\eta(0,\slashed{D})$ to the anomaly in some effective boundary theory. We have to stress however that these arguments work for strongly elliptic boundary conditions only.

The local part of the parity anomaly on a manifold without boundary for gauge and gravitational theories is given by a Chern--Simons action, see \cite{Witten:1985xe} for a discussion on this correspondence.  

Many authors use the so-called APS $\eta$-invariant instead of $\eta(0,\slashed{D})$, which is defined as
\begin{equation}
\eta_{\slashed{D}}=\frac 12 \bigl(\eta(0,\slashed{D}) +\mathrm{dim}\, \mathrm{Ker}\, \slashed{D} \bigr),\label{etainv}
\end{equation}
when $\eta(s,\slashed{D})$ is regular at $s=0$. This invariant behaves somewhat more regularly, when an eigenvalue of $\slashed{D}$ crosses the origin. Namely, $\eta_{\slashed{D}}$ is continuous when $\lambda\to +0$. The exponentiated $\eta$-invariant
\begin{equation}
\mathcal{E}(\slashed{D})=\exp(-2\pi\ii\eta_{\slashed{D}}) \label{expeta}
\end{equation}
is even a smooth function under variations of the background fields. 

If $\slashed{D}$ is the Dirac Hamiltonian, then $\eta(0,\slashed{D})$ describes the fermion number fractionization due to the Jackiw--Rebbi mechanism \cite{Jackiw:1975fn}, see  \cite{Niemi:1984vz} for review.

Let us now describe the plan of this work and the main results. In the next section, we analyse Euclidean bag boundary conditions and extend the previous works \cite{Kurkov:2017cdz,Kurkov:2018pjw} to arbitrary even-dimensional manifolds. We use a two-step procedure. Firstly, basing on general properties of the heat kernel expansion, we identify the form of $\eta(0,\slashed{D})$ up to a universal function of a boundary Dirac operator. Secondly, this universal function is determined through an exactly solvable example. Thus, the $\eta$-invariant of $\slashed{D}$ is expressed through $\eta$-invariants of boundary Dirac operators. The rest of the work is dedicated to odd-dimensional manifolds. In Section \ref{sec:str}, we analyse the Witten--Yonekura boundary conditions, which are valid in any number of dimensions. Although the anomaly inflow for these conditions has been already analysed in \cite{Witten:2019bou} it is tempting to reobtain the results from the heat kernel point of view (since the heat kernel is usually very effective for the anomaly calculations). As we have mentioned above, to ensure applicability of our methods, boundary conditions have to satisfy the strong ellipticity condition. We show, that WY boundary conditions are not strongly elliptic, but very close to strong ellipticity. As a by-product, we analyse strong ellipticity of complexified chiral bag boundary conditions. Finally, in Section \ref{sec:odd}, we study a family of strongly elliptic local boundary conditions on odd-dimensional manifolds. By using roughly the same strategy as in Section \ref{bag}, we show that the $\eta$-invariant of $\slashed{D}$ is given by the indices of boundary Dirac operators. Possible extensions of our method are briefly described in Section \ref{sec:con}.

\section{Euclidean bag boundary conditions}\label{bag}
Bag boundary conditions were introduced in the context of hadron physics in \cite{Chodos:1974je,Chodos:1974pn}. The characteristic property of these boundary conditions, besides the locality, is that the normal component of fermion current vanishes on the boundary, so that quarks cannot escape the bag. This also makes the Euclidean Dirac operator hermitian. We fix the chirality matrix $\gamma_*$ to be
\begin{equation}
\gamma_*=-\frac{\ii^{n/2}}{n!}\, \epsilon^{\mu_1\mu_2\dots \mu_n}\gamma_{\mu_1}\gamma_{\mu_2}\dots\gamma_{\mu_n}. \label{chir}
\end{equation}
Let $\nn$ be an inward pointing unit normal to the boundary, and let $x^a$ denote coordinates on the boundary. Let the boundary $\partial\MM$ contain several connected components $\partial\MM_\alpha$. We take the boundary projector in the form
\begin{equation}
\Pi_-=\tfrac 12 (1-\ii\varepsilon_\alpha\gamma_*\gamma^\nn ), \label{Pimin}
\end{equation}
where the sign factor $\varepsilon_\alpha=\pm 1$ is constant on each $\partial\MM_\alpha$.
It is convenient to introduce a complementary projector $\Pi_+=\tfrac 12 (1+\ii\varepsilon_\alpha\gamma_*\gamma^\nn )$ and $\chi\equiv \Pi_+-\Pi_-=\ii\varepsilon_\alpha\gamma_*\gamma^\nn$.

Condition (\ref{bc}) defines a strongly elliptic boundary value problem with $\eta(0,\slashed{D})$ being regular at $s=0$, see \cite{Gilkey:1983xz}. The first two heat kernel coefficients read
\begin{eqnarray}
&&a_0(Q,\slashed{D}^2)=\frac 1{(4\pi)^{n/2}}\int_{\MM}d^n x\, \sqrt{g}\, \mathrm{tr}\, (Q),\label{a0}\\
&&a_1(Q,\slashed{D}^2)=\frac 1{4(4\pi)^{(n-1)/2}}\int_{\partial\MM}d^{n-1} x\, \sqrt{h}\, \mathrm{tr}\, (Q\chi ),\label{a1}
\end{eqnarray}
where $g$ and $h$ are the determinants of the metric tensors on the bulk and the boundary, respectively. Let us additionally note that for a boundary Dirac operator $\mathcal{D}$ defined below, the first heat kernel coefficient is expressed in the same manner
as in (\ref{a0})
\begin{equation}
a_0(Q,\mathcal{D}^2)=\frac 1{(4\pi)^{(n-1)/2}}\int_{\partial\MM}d^{n-1} x\, \sqrt{h}\, \mathrm{tr}\, (Q).\label{a000}
\end{equation}

We assume that each component of the boundary $\partial\MM_\alpha$ has a collar neighbourhood, where $\MM$ has a direct product structure $\partial\MM_\alpha\times [0,\tau_\alpha )$ with some positive real number $\tau_\alpha$. Let $x^\nn\in [0,\tau_\alpha )$ be a normal coordinate, which is equal to the geodesic distance to the boundary. Then $(x^a,x^\nn )$ becomes a coordinate system in this neighbourhood, such that $g^{\nn\nn}=1$ and $g^{a\nn}=0$. The product structure implies that $g_{ab}$ and $A_a$ do not depend on $x^\nn$. The component $A_\nn$ may be removed by a gauge transformation. Let us take the $\gamma$-matrices in the form
\begin{equation}
\gamma^a=\begin{pmatrix}
\Gamma^a & 0 \\ 0 & -\Gamma^a 
\end{pmatrix},\qquad
\gamma^\nn=\begin{pmatrix}
0 & 1 \\ 1 & 0
\end{pmatrix},
\qquad
\gamma_* =\begin{pmatrix}
0 & \ii \\ -\ii & 0
\end{pmatrix}. \label{chirmatm}
\end{equation}
We have chosen one of two inequivalent representation of the Clifford algebra on the boundary generated by $\Gamma^a$ by imposing the relation
\begin{equation}
\epsilon^{a_1\dots a_{n-1}\nn}\Gamma_{a_1}\dots \Gamma_{a_{n-1}}=(n-1)!(-\ii)^{(n+2)/2}.
\label{Gams}
\end{equation}
The other representation corresponds to a minus sign on the right hand side. Since in even-dimensional Euclidean spaces the Clifford algebra has a unique irreducible representation up to the equivalence, the choice of sign in (\ref{Gams}) does not play any role. 

The Dirac operator near the boundary reads 
\begin{equation}
\slashed{D}=\begin{pmatrix}
\mathcal{D} & \ii \partial_\nn \\ \ii \partial_\nn & -\mathcal{D}
\end{pmatrix}, \label{Dirm}
\end{equation}
where $\mathcal{D}=\ii \Gamma^a\nabla_a$. In accordance to the block decompositions of the Dirac operator we also decompose the spinor field as  $\psi =(\varphi_1,\varphi_2)^T$. For $\varepsilon_\alpha=1$ the boundary condition is $\varphi_1\vert_{\partial\MM_\alpha}=0$, while for $\varepsilon_\alpha=-1$ one has $\psi_2\vert_{\partial\MM_\alpha}=0$. Thus, non-zero boundary values of $\psi$ are given by $\varphi_2$ and $\varphi_1$, respectively. One can easily obtain that the restriction of $\slashed{D}$ on the space of boundary values is 
\begin{equation}
\mathcal{D}_\alpha:=-\varepsilon_\alpha \mathcal{D}.\label{Dalpha}
\end{equation}

As we have explained above, see Section \ref{sec:in}, the variations of $\eta(0,\slashed{D})$ are expressed through odd-numbered heat kernel coefficients $a_{n-1}$ and $a_{n+1}$, which do not contain any bulk contributions. The boundary contributions are integrals of local invariants associated with $\slashed{D}$. They are constructed by using the Riemann tensor, the extrinsic curvature, the field strength, and their derivatives (both normal and tangential). Due to our restriction to the field configurations having a product structure near the boundary, only the invariants depending on the boundary metric, the gauge field $A_a$, and their tangential derivatives survive. Such invariants can be associated with the operator $\mathcal{D}$. The heat kernel coefficients depend also on the parameters $\varepsilon_\alpha$ though this dependence is easy to control. Let us invert the sign $\varepsilon_\alpha\to -\varepsilon_\alpha$. In a small neighbourhood of the boundary this just means exchanging the roles of $\varphi_1$ and $\varphi_2$ and thus is equivalent to $\mathcal{D}\to -\mathcal{D}$. Therefore, the heat kernel coefficients may depend on $\varepsilon_\alpha$ and $\mathcal{D}$ only through their product $\mathcal{D}_\alpha$. The same can be said about $\eta(0,\slashed{D})$. I.e., there is a function $f$ such that
\begin{equation}
\eta(0,\slashed{D})=\sum_\alpha f(\mathcal{D}_\alpha ).\label{etaf}
\end{equation}

Our task is to define the function $f$. Below, we will demonstrate that
\begin{equation}
f(\mathcal{D}_\alpha)=\tfrac 12 \eta(0,\mathcal{D}_\alpha).\label{fD}
\end{equation}

We stress that equation (\ref{etaf}) means that smooth variations of both sides coincide. A better result can be formulated for the exponentiated $\eta$-invariant (\ref{expeta}), which is smooth and allows for integration of smooth variations
\begin{equation}
\mathcal{E}^2(\slashed{D})=C \prod_\alpha \mathcal{E}(\mathcal{D}_\alpha),\label{Ebag}
\end{equation}
where $C$ is an integration constant, i.e. it is a topological functional with vanishing local variations.
\subsection{Two-dimensional manifolds}\label{sec:two}
As a warm-up, we present here an explicit calculation of variations of the $\eta$ functions for $n=2$. We take the tangent one-dimensional gamma matrix as $\Gamma=1$. The boundary $\partial\MM$ is diffeomorphic to a disjoint unit of spheres $S^1$ with unit metrics. Thus, due to the product structure assumption the geometry is flat near the boundaries. The variation of $\eta(0,\slashed{D})$ contains boundary contributions only. Thus, it is enough to consider the variation of the gauge connection, so that
\begin{equation}
\delta\slashed{D}=\delta A_\sigma \begin{pmatrix}
\ii & 0 \\ 0 & -\ii
\end{pmatrix}. \label{Qn2}
\end{equation}  
By substituting this expression in (\ref{vareta}), using (\ref{a1}), and computing the trace over spinor indices, one obtains
\begin{equation}
\delta\eta(0,\slashed{D})=\frac {\ii}{2\pi} \sum_{\alpha}\varepsilon_\alpha \int_{\partial\MM_\alpha}\dd \sigma\, \mathrm{tr}_G\, \delta A_\sigma , \label{var1n2}
\end{equation}
where the trace $\mathrm{tr}_G$ extends to the gauge indices only. The coordinate on $S^1$ has been denoted by $\sigma$.

Let us compute the variation of $\eta(0,\mathcal{D})$ for the boundary operator $\mathcal{D}$ on a component $\partial\MM_\alpha$ fixed. We have
\begin{equation}
\mathcal{D}=\ii (\partial_\sigma + A_\sigma ),\qquad \delta \mathcal{D}=\ii \delta A_\sigma.
\end{equation}
Then, using (\ref{a000}), we get
\begin{equation}
\delta\eta (0,\mathcal{D})= -\frac{\ii}{\pi} \int_{\partial\MM_\alpha}\dd \sigma\, \mathrm{tr}_G\, \delta A_\sigma. \label{var2in2}
\end{equation}
This result is consistent with (\ref{etaf}) and (\ref{fD}) if one takes into account (\ref{Dalpha}) and makes the summation by $\alpha$.

We conclude this subsection with a discussion of the gauge invariance.
Equations (\ref{var1n2}) and (\ref{var2in2}) describe all variations of the corresponding $\eta$ functions except for the case, when an eigenvalue of $\slashed{D}$ or $\mathcal{D}$ crosses the origin and thus $\eta(0,\slashed{D})$ or $\eta(0,\mathcal{D})$ jumps by 2. Thus, we may expect that these variations vanish modulo $2\mathbb{Z}$ for a gauge transformation. This is a rather nontrivial consistency check. Note that (\ref{var1n2}) and (\ref{var2in2}) contain only the traces of the gauge fields. Thus, it is sufficient to consider just an abelian  gauge group with gauge transformations, $\delta_\phi A_\sigma=\ii \partial_\sigma \phi$ with $\phi$ being a gauge parameter. Such transformations are linear. There is no need to restrict our attention to infinitesimal values of $\phi$. The gauge group element $e^{i\phi}$ has to be periodic on each $S^1=\partial\MM_\alpha$. Therefore, $\phi$ has to be quasiperiodic, $\phi(\sigma+2\pi)=\phi(\sigma)+2\pi N_\alpha$, $N_\alpha\in\mathbb{Z}$, which yields the required gauge-invariance property of $\eta(0,\mathcal{D})$
\begin{equation}
\delta_\phi \eta (0,\mathcal{D}) =2 \sum_\alpha N_\alpha = 0\ \mbox{mod}\ 2\mathbb{Z}.
\end{equation}

This equation is not yet enough to demonstrate the gauge invariance of $\eta(0,\slashed{D})$ modulo $2\mathbb{Z}$. A stronger relation may be derived with the help of usual homology theory arguments. Let us split $\MM$ into a union of a finite number of simply connected plaquettes $\mathcal{P}_k$. The plaquettes inherit their orientations from $\MM$. Two neighbouring plaquettes intersect on their common boundaries. Since plaquettes are simply connected, their boundaries are contractible, and 
\begin{equation}
\int_{\partial\mathcal{P}_k}\dd \phi =0.\label{intPk}
\end{equation}
Let us sum up Eq.\ (\ref{intPk}) over $k$. The contributions of pieces of the boundaries of plaquettes which belong to the interior of $\MM$ cancel since these pieces enter the sum twice but with opposite orientations. We are left with
\begin{equation}
\frac 1{2\pi}\sum_{\alpha} \int_{\partial\MM_\alpha}\dd \sigma\, \partial_\sigma \phi=\frac 1{2\pi}\sum_k \int_{\partial\mathcal{P}_k} \dd\phi =0.
\label{sumPk}
\end{equation}
Each term in the sum over $\alpha$ in (\ref{sumPk}) is an integer. If one flips the sign in front of a term, the whole sum thus changes by an even integer. Therefore, we obtain
\begin{equation}
\frac 1{2\pi}\sum_{\alpha} \varepsilon_\alpha \int_{\partial\MM_\alpha}\dd \sigma\, \partial_\sigma \phi =0 \ \mbox{mod}\ 2\mathbb{Z}.
\end{equation}
This proves that
\begin{equation}
\delta_\phi\eta(0,\slashed{D})=0 \ \mbox{mod}\ 2\mathbb{Z}.
\end{equation}
\subsection{General even dimension}\label{sec:gen}
To define the function $f$ in Eq.\ (\ref{etaf}) for any even dimension $n$, it is sufficient to calculate the $\eta(0,\slashed{D})$ for a given boundary and a given $\mathcal{D}$ with any suitable choice of the bulk geometry. The simplest case is a cylinder with two identical boundaries, $\MM = \mathcal{N}\times [0,\pi ]$. Practically, we repeat below an example by Gilkey and Smith \cite{Gilkey:1983a} with some small modifications. The coordinate $\tau\in [0,\pi]$ coincides with $x^\nn$ at $\tau=0$. We extend the $\gamma$ matrices defined near $\tau=0$ in (\ref{chirmatm}) to the whole manifold
\begin{equation}
\gamma^a=\begin{pmatrix}
\hat\Gamma^a & 0 \\ 0 & -\hat\Gamma^a 
\end{pmatrix},\qquad
\gamma^\tau=\begin{pmatrix}
0 & 1 \\ 1 & 0
\end{pmatrix}. \label{newGam}
\end{equation}
It is assumed that the matrices $\hat\Gamma^a$ satisfy (\ref{Gams}), so that $\gamma_*$ remains as in (\ref{chirmatm}). We introduced new notations, since near $\tau=\pi$ one has $\gamma^\nn=-\gamma^\tau$ and $\Gamma^a=-\hat\Gamma^a$. The Dirac operator reads
\begin{equation}
\slashed{D}=\begin{pmatrix}
\hat{\mathcal{D}} & \ii \partial_\tau \\ \ii \partial_\tau & -\hat{\mathcal{D}}
\end{pmatrix}, \label{newDir}
\end{equation}
where $\hat{\mathcal{D}}=\ii \hat\Gamma^a\nabla_a$. The spectrum of $\slashed{D}$ can be found in terms of the spectrum of $\hat{\mathcal{D}}$. So we use the relation $\hat{\mathcal{D}}\varphi(\kappa)=\kappa\varphi(\kappa)$.

Let us take $\varepsilon\vert_{\tau=0}=-1$, which means $\varphi_2\vert_{\tau=0}=0$. Then the operator (\ref{Dalpha}) is 
\begin{equation}
\mathcal{D}\vert_{\tau=0}=\hat{\mathcal{D}}.
\end{equation}
With the choice $\varepsilon\vert_{\tau=\pi}=1$, we have $\varphi_2\vert_{\tau=\pi}=0$, and thus $\mathcal{D}\vert_{\tau=\pi}=\hat{\mathcal{D}}$. This problem has two families of eigenfunctions. The first one is
\begin{eqnarray}
&&\varphi_2= \sin (k\tau) \varphi(\kappa),\\
&&\varphi_1=\frac {\ii}k \, \cos(k\tau) (\lambda + \kappa)\varphi(\kappa),\label{spec1}
\end{eqnarray}
with $k$ being a positive integer. Allowed eigenvalues are $\lambda=\pm \sqrt{k^2+\kappa^2}$. This part of the spectrum is symmetric and thus does not contribute to the $\eta$ function. The second family corresponds to
\begin{equation}
\varphi_1=\varphi(\kappa),\qquad \varphi_2=0,\qquad \lambda=\kappa.
\end{equation}
Therefore, we conclude that
\begin{equation}
\eta(s,\slashed{D})=\eta(s,\hat{\mathcal{D}}).\label{etaeta}
\end{equation}
Using Eq.\ (\ref{etaf}), we obtain $\eta(0,\slashed{D})=f(\mathcal{D}\vert_{\tau=0})+f(\mathcal{D}\vert_{\tau=\pi})=2f(\hat{\mathcal{D}})$. This proves Eq.\ (\ref{fD}) announced above.

As a consistency check, we also analyse the case $\varepsilon\vert_{\tau=\pi}=-1$, $\varphi_1\vert_{\tau=\pi}=0$. The spectral resolution consists of the modes (\ref{spec1}) with $k\in \mathbb{N}_0+\tfrac 12$ with symmetric eigenvalues, so that
\begin{equation}
\eta(s,\slashed{D})=0.\label{etaeq0}
\end{equation}
In this case, $\mathcal{D}\vert_{\tau=0}=-\mathcal{D}\vert_{\tau=\pi}=\hat{\mathcal{D}}$, so that (\ref{etaeq0}) is consistent with (\ref{etaf}) and (\ref{fD}).

\section{Other boundary conditions and strong ellipticity}\label{sec:str}
\subsection{Definitions}\label{sec:def}
In our approach to a strong ellipticity we mostly follow \cite{Gilkey:1983a,Gilkey:1983xz,GilkeyNew}, where the reader may find more details and proofs.
The strong ellipticity of a spectral problem is defined with respect to a cone $\mathcal{C}$ in the complex plane, which is a subset in $\mathbb{C}$ invariant under multiplications by non-negative real numbers. There is a family of cones 
\begin{equation}\label{str2}
\mathcal{C}_{T}=\{\lambda\in\mathbb{C}:|{\Im}(\lambda)|\geqslant
\tan(\pi T/4)
|{\Re}(\lambda)|\},\,\,\,\mbox{where}\,\,\,
T\in(0,1],
\end{equation}
which plays a particularly important role.

To obtain the leading symbol $p(x,k)$ of a differential operator, one has to take the terms with highest-order derivatives and replace there each partial derivative by $-\ii k$ with the wave vector $k\in T^*\MM$ at each point of $\MM$. For $\slashed{D}$, one has $p(x,k)=\gamma^\mu k_\mu$. The strong ellipticity of the operator itself means that
\begin{equation}\label{str4}
	\det\big(p(x,k)-\lambda\big)\neq 0\,\,\,\mbox{for all}\,\,\,(k,\lambda)
	\in T^*\mathcal{M}\times\mathcal{C}, \quad (k,\lambda)\neq(0,0).
	\end{equation}
For the Dirac operator, this condition is trivially satisfied for $\mathcal{C}=\mathcal{C}_T$ with $T\in (0,1]$. 

In some sense, the leading symbol of an operator defines the main part of this operator in the ultraviolet regime. To analyse strong ellipticity of a boundary value problem one has to define some main part of this problem. Let $\xi\in T^*\partial\MM$ be a boundary wave vector, so that $k=(\xi,0)$. Let
\begin{equation}\label{str1}
q(\xi,\lambda)=-\ii\gamma^\nn\big(p((y,0),(\xi,0))-\lambda\big)=-\ii\gamma^\nn\bigl(\gamma^a((y,0))\xi_a-\lambda\bigr).
\end{equation}

\begin{definition}\label{def2}
We say that the boundary value problem for $\slashed{D}$ with the boundary projector $\Pi_-$ is strongly elliptic with respect to $\mathcal{C}$ iff for any $(\xi,\lambda)
		\in T^*(\partial\mathcal{M}_\alpha)\times\mathcal{C}$, $(\xi,\lambda)\neq(0,0)$, there is no solution $g(r)$, $r\in \mathbb{R}_+$, for the auxiliary boundary value problem 
		\begin{equation}\label{str15}
\big(\partial_r+q(\xi,\lambda)\big)g(r)=0,\,\,\,\Pi_-g(0)=0,\,\,\,
\lim_{r\to+\infty}g(r)= 0.
\end{equation}
\end{definition}

Roughly speaking, strong ellipticity means that at the ultraviolet region the main bulk and boundary parts of the spectral problem for $\slashed{D}$ are not allowed to have eigenvalues in the "wrong part" of the complex plane. There is a hierarchy of strong ellipticity conditions depending on the cone $\mathcal{C}$. The weakest condition corresponds to the smallest $\mathcal{C}=\{ 0\}$. In this case, one only has the elliptic smoothness property: generalized solutions of the boundary value problem are actually smooth. Such boundary conditions are usually called elliptic (without "strongly") in the literature. Strong ellipticity with respect to the cone $\mathcal{C}_1$ is the minimal requirement to ensure existence of the $\eta$ function, locality of the heat kernel, etc. I.e., this condition is necessary for applicability of our methods. The point $T=0$ is not included in the definition (\ref{str2}). We define
\begin{equation}
\mathcal{C}_0\equiv \cup_{T\in (0,1]}\, \mathcal{C}_T=\mathbb{C}-(0,+\infty)-(-\infty,0).
\label{C0}
\end{equation}
Strong ellipticity with respect to $\mathcal{C}_0$ implies that the spectrum of auxiliary value problem is real, which brings up some advantages, see \cite{GilkeyNew}.

If there are no boundaries, (strong) ellipticity guarantees the existence of the DeWitt expansion for the heat kernel with the main term defined by the principal symbol. Similar ideas apply in the case with boundaries \cite{GilkeyNew}, see also \cite{McAvity:1990we} where the DeWitt expansion with local boundary conditions was discussed.

\subsection{Witten--Yonekura boundary conditions}\label{sec:WY}
Let us consider the boundary conditions defined by 
\begin{equation}\label{str19}
\Pi_-=\frac{1}{2}
\big(1-\varepsilon_\alpha\gamma^\nn\big),\,\,\,
\mbox{where}\,\,\,\varepsilon_\alpha\in\{-1,1\}.
\end{equation}
To the best of our knowledge, for the first time such boundary conditions appeared on Euclidean manifolds in \cite{Witten:2015aba}, while the anomaly inflow was studied in \cite{Witten:2019bou}. The normal component of the fermion current $\psi^\dag \gamma^\nn \psi$ does not vanish on $\partial\MM$ for the boundary conditions defined by (\ref{str19}). Therefore, $\slashed{D}$ is not hermitian and $\eta(0,\slashed{D})$ does not define the phase of $\det \slashed{D}$. The work \cite{Witten:2019bou} used a Pauli--Villars type expression for the fermion path integral and thus avoided this problem. Thus a natural question arises: Is it possible to use the heat kernel methods to achieve an alternative understanding of the results of \cite{Witten:2019bou}? The first thing to check is whether the boundary conditions (\ref{str19}) are strongly elliptic.

It appears, that it is sufficient to consider $\xi=0$.
Let us fix $\alpha$ and $\lambda\in\mathcal{C}_{T}$, such that $T\in(0,1]$ and ${\Im}(\varepsilon_\alpha\lambda)<0$. Then $q(0,\lambda)=\ii\gamma^\nn\lambda$ and hence the problem (\ref{str15}) reads
\begin{equation}\label{str20}
\big(\partial_r+\ii\gamma^\nn\lambda\big)g(r)=0,\,\,\,\big(1-\varepsilon_\alpha\gamma^\nn\big)g(0)=0,\,\,\,\lim_{r\to+\infty}g(r)= 0.
\end{equation}
It is quite easy to see that we can take any vector $g_0$ from the kernel of $\Pi_-$, and write the solution in the form
\begin{equation}\label{str21}
g(r)=g_0e^{-\varepsilon_\alpha \ii\lambda r}.
\end{equation}
This means, that the conditions of Definition \ref{def2} are not satisfied. Hence, the spectral problem with the boundary conditions defined in (\ref{str19}) is not strongly elliptic with respect to $\mathcal{C}_T$ for $T\in [0,1]$.

Some comments and clarifications are in order.
\begin{enumerate}
\item One can check that WY boundary conditions are strongly elliptic with respect to $\mathcal{C}=\{ 0\}$ as has been already stated in \cite{Witten:2019bou}. Thus the boundary value problem is elliptic in the usual classification and the Dirac operator admits a parametrix (or quasi inverse), which is necessary for constructing a propagator and Feynman diagrams. At the same time, the resulting QFT has some exotic properties. For example, the partition function cannot be expressed through a product of eigenvalues. This property is in correspondence with the properties of the desired effective boundary theory, the chiral fermions, for which the path integral is also not a product of eigenvalues. In the view of these properties, the loss of strong ellipticity may be not so surprising. An interesting question arises: Does strong ellipticity condition exclude chiral boundary fermions? We do not know the answer at the moment.
\item To demonstrate the absence of strong ellipticity it was enough to consider $\xi=0$. It can be shown that the problems appear for vanishing $\xi$ only. Therefore, if $\xi=0$ is excluded (e.g. by imposing an anti-periodicity condition with respect to one of the boundary coordinates) at least some of the limitations imposed by the absence of the strong ellipticity may be removed. However, there are no general statements in this case.
\item In Section \ref{sec:comp}, we show that WY boundary conditions on even-dimensional manifolds are infinitely close to strongly elliptic boundary conditions.
\end{enumerate}

\subsection{Complexified chiral bag boundary conditions}\label{sec:comp}
Let $n$ be even. Consider the boundary projector
\begin{equation}\label{str17}
\Pi_-(\theta)=\frac{1}{2}
\bigl(1-\ii\varepsilon_\alpha {\gamma_*}e^{\theta{\gamma_*}}\gamma^\nn\bigr),\,\,\,
\varepsilon_\alpha\in\{-1,1\}.
\end{equation}
It defines the so-called chiral bag boundary conditions \cite{Rho:1983bh}. Chiral and parity anomalies for these boundary conditions with a real $\theta$ were considered in \cite{Hrasko:1983sj,Wipf:1994dy,Gilkey:2005qm,Kirchberg:2006wu,Ivanov:2021yms}.
The strong ellipticity of these boundary conditions with respect to the cone $\mathcal{C}_0$  for real values of $\theta$ was demonstrated in \cite{Beneventano:2003hv}. Here we extend this analysis for complex chiral angles $\theta$. The values $\theta=\pm\ii \pi/2$ correspond to WY boundary conditions. Since the strong ellipticity with respect to $\mathcal{C}_1$ represent minimal requirements for the existence of main spectral functions, we perform our analysis in this cone.

It is convenient to take the $\gamma$-matrices in a slightly different form than in (\ref{chirmatm}):
\begin{equation}\label{str5}
\gamma^a=\begin{pmatrix}
0& \ii\\
-\ii& 0
\end{pmatrix}\otimes{\Gamma}^a,\,\,\,
\gamma^\nn=\begin{pmatrix}
0& 1\\
1& 0
\end{pmatrix}\otimes\mathrm{id},\,\,\,
\gamma_*=\begin{pmatrix}
1& 0\\
0& -1
\end{pmatrix}\otimes\mathrm{id}.
\end{equation}
We restrict ourselves to the case $\varepsilon_\alpha=1$ since $\varepsilon_\alpha=-1$ can be reached by the shift $\theta\to \theta+\ii\pi$. In the basis (\ref{str5}), we have
\begin{equation}\label{str9}
\Pi_-(\theta)=\frac{1}{2}\begin{pmatrix}
1&-\ii e^{\theta}\\
\ii e^{-\theta}&1
\end{pmatrix}\otimes\mathrm{id}.
\end{equation}
Thus, the kernel of boundary projector reads
\begin{equation}\label{str10}
\mathrm{Ker}\big(\Pi_-(\theta)\big)=\bigg\{
\begin{pmatrix}
-1\\
\ii e^{-\theta}
\end{pmatrix}\otimes v:\,v\in \mathcal{V}\bigg\},
\end{equation}
where $\mathcal{V}$ denotes the space of boundary spinors.

The matrix (\ref{str1}) gets the form
\begin{equation}\label{str6}
q(\xi,\lambda)=
\begin{pmatrix}
-{\Gamma}^a\xi_a& \ii\lambda\\
\ii\lambda& {\Gamma}^a\xi_a
\end{pmatrix}.
\end{equation}
Solutions of the first equation in (\ref{str15}) vanishing at $r\to+\infty$ correspond to $g(0)$ belonging to the space $V_+(\xi,\lambda)$, which is spanned by the eigenvectors of $q(\xi,\lambda)$ corresponding to eigenvalues with a positive real part. 

Let us analyse the case $\xi\neq 0$. The space $\mathcal{V}$ splits into a direct sum of spaces $\mathcal{V}_\pm$ corresponding to positive (negative) eigenvalues of $\Gamma^a\xi_a$, respectively. Namely, for $v_\pm\in \mathcal{V}_\pm$ one has $\Gamma^a\xi_a v_\pm=\pm |\xi|v_\pm$. Then, $V_+(\xi,\lambda)$ is spanned by the vectors
\begin{equation}\label{str7}
\begin{pmatrix}
\mp |\xi|+\sqrt{|\xi|^2-\lambda^2}\,\\
\ii \lambda
\end{pmatrix}\otimes v_{\pm}:\ v_{\pm}\in \mathcal{V}_\pm(\xi).
\end{equation}
The corresponding eigenvalue of $q$ is $\sqrt{|\xi|^2-\lambda^2}$. It has a positive real part for $\lambda\in\mathcal{C}_1$. Boundary conditions are strongly elliptic iff
\begin{equation}\label{str11}
V_{+}(\xi,\lambda)\cap\mathrm{Ker}\big(\Pi_-(\theta)\big)=\{0\}.
\end{equation}
Since $v_+$ and $v_-$ are linearly independent, the condition (\ref{str11}) means that the 2-vectors 
\begin{equation}
\begin{pmatrix}
\mp |\xi|+\sqrt{|\xi|^2-\lambda^2}\,\\
\ii \lambda
\end{pmatrix}\quad \mbox{and} \quad \begin{pmatrix}
-1\\
\ii e^{-\theta}
\end{pmatrix}  \label{str2vect}
\end{equation}
can never be proportional to each other. This in turn implies that the strong ellipticity is equivalent to the statement that the equation
\begin{equation}
\lambda e^{\theta}=-\bigl(\mp |\xi| +\sqrt{|\xi|^2 -\lambda^2} \bigr) \label{streq1}
\end{equation}
has no solutions. Since $\mathcal{C}_1$ is invariant under the reflections $\lambda\to -\lambda$, this condition can be transformed to the condition of absence of solutions for
\begin{equation}
\lambda e^{2\theta}+2|\xi|e^\theta +\lambda=0,\label{streq2}
\end{equation}
which is equivalent to
\begin{equation}
\lambda =-\frac{|\xi|}{\mathrm{cosh}\,(\theta)}.\label{streq3}
\end{equation}
This equation has no solutions with $\lambda\in\mathcal{C}_1$ if
\begin{equation}\label{str12}
\arg\big(\cosh(\theta)\big)\in \bigg( -\frac{\pi}4,\frac{\pi} 4\bigg) \cup \bigg( \frac{3\pi}4,\frac{5\pi}4\bigg).
\end{equation}
Analysis of the case $\xi=0$ is elementary and does not alter the result.

\begin{figure}[h]
	\centerline{\includegraphics[width=0.6\linewidth]{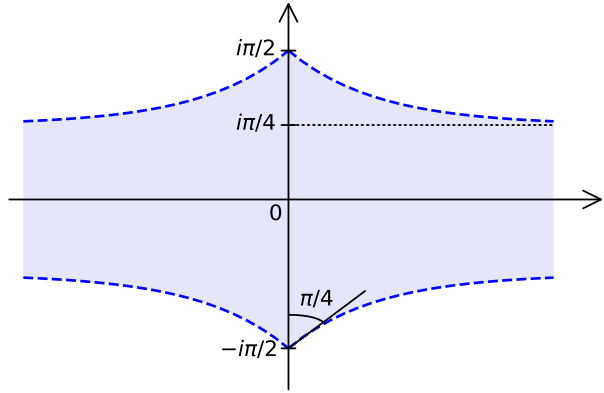}}
	\caption{The painted over domain of complex plane of $\theta$ is $\Theta$. It does not include the boundary.}
	\label{str3}
\end{figure}

We conclude that the strongly elliptic boundary conditions correspond to the values 
\begin{equation}
\theta\in \{ \Theta +\ii \pi N\},\quad N\in\mathbb{Z}, \label{strdom}
\end{equation}
with the fundamental domain $\Theta$ depicted at Fig.\ \ref{str3}. Actually, the chiral angle $\theta$ may even depend on the boundary points. The boundary value problem is strongly elliptic as long as $\theta$ belongs everywhere to the set (\ref{strdom}).

As we have mentioned above, the WY boundary conditions correspond to $\theta=\pm \ii\pi/2$. A perturbation of this value by an arbitrarily small imaginary number leads to strongly elliptic boundary conditions.

\section{Anomaly inflow in odd dimensions}\label{sec:odd}
Boundary conditions for a Dirac operator are called Clifford boundary conditions if $\chi \gamma^\nn = -\gamma^\nn \chi$ and $\chi \gamma^a=\gamma^a\chi$. It is known \cite{Gilkey:1983xz}, that such boundary conditions ensure strong ellipticity of the boundary value problem and the regularity of $\eta(s,\slashed{D})$ at $s=0$. Besides, the Dirac operator with Clifford boundary conditions is Hermitian. Euclidean bag boundary conditions (\ref{bag}) in even dimensions provide an example of boundary conditions belonging to this class. If $n$ is odd, such boundary conditions do not exist for a single generation of fermions. Thus, we double the number of fermions and take the $\gamma$ matrices in the form 
\begin{equation}\gamma^\mu = \Gamma^\mu \otimes \sigma_2, 
\end{equation}
where $\sigma_i$ for $i=1,2,3$ denotes the standard Pauli matrix, $\Gamma^\mu$ is an irreducible representation of Dirac matrices on $n$ dimensions tensored with an identity matrix in the gauge space. Such representations of $\gamma$ matrices appear for fermions reduced from $n+1$ to $n$ dimensions and when passing from an $(n+1)$-dimensional Minkowski theory to corresponding Dirac hamiltonian in $n$ dimensions. If $\partial\MM=\emptyset$, we have two identical copies of a fermionic theories related by the parity inversion $\Gamma^\mu\to -\Gamma^\mu$. Therefore, the parity anomaly vanishes. Another way to see this is to realize that the matrix 
\begin{equation}
\mathrm{id}\otimes \sigma_3 \label{anti}
\end{equation}
anticommutes with $\slashed{D}$ and thus the spectrum is symmetric. These arguments break down in the presence of a boundary if the boundary conditions do not commute with (\ref{anti}).

\subsection{Particular case calculations}\label{sec:part}
As in Section \ref{sec:gen}, we take $\mathcal{M}=\mathcal{N}\times[0,\pi]$. We define the family of projectors
\begin{equation}\label{ps2}
\Pi_-(\vartheta)=\frac{1}{2}\big(1-\Gamma^\tau \otimes(\cos(2\pi\vartheta)\sigma_3+\sin(2\pi\vartheta)\sigma_1)\big),
\end{equation}
parametrized by a real angle $\vartheta$. The matrix $\mathrm{id}\otimes (\cos(2\pi\vartheta)\sigma_3+\sin(2\pi\vartheta)\sigma_1)$ commutes with $\Pi_-$ and anticommutes with $\slashed{D}$. Therefore, two identical boundary projectors for $\tau=0$ and $\tau=\pi$ lead to a symmetric non-zero spectrum of the Dirac operator and thus to a vanishing $\eta$ function. For our purposes, it is sufficient to consider $\theta$ being non-zero on a single component of the boundary
\begin{equation}\label{ps3}
\Pi_-(\vartheta)\psi\big|_{\tau=0}=0,\qquad
\Pi_-(0)\psi\big|_{\tau=\pi}=0.
\end{equation} 
Obviously, the spectral problem is periodic with respect to $\vartheta\to\vartheta+1$, and at $\vartheta=0$ the $\eta$ function vanishes. In what follows, we consider $\vartheta\in [0,1)$.

The Dirac operator reads
\begin{equation}\label{ps1}
\slashed{D}=\ii\Gamma^{\mu}\nabla_{\mu}\otimes\sigma_2=
\mathcal{D}\otimes\sigma_2+\ii\Gamma^\tau\partial_\tau \otimes\sigma_2,
\,\,\,\mbox{where}\,\,\,\,\mathcal{D}=i\Gamma^{a}\nabla_{a}.
\end{equation}
As in Section \ref{sec:gen}, the operator admits separation of variables, so that the spectral problem $\slashed{D}\psi_\lambda =\lambda\psi_\lambda$ becomes essentially algebraic. The matrix $\Gamma^\tau$ anticommutes with all $\Gamma^a$ and thus plays the role of a chirality operator on $\mathcal{N}$.

We start with non-zero eigenvalues of $\mathcal{D}$. Since $\Gamma^\tau\mathcal{D}=-\mathcal{D}\Gamma^\tau$, the corresponding eigenfunctions split in sums of opposite chirality parts, $\varphi^{\pm}(\kappa)$, such that $\mathcal{D}\varphi^{\pm}(\kappa)=\kappa\varphi^{\mp}(\kappa)$ and $\Gamma^\tau \varphi^{\pm}(\kappa)=\pm \varphi^{\pm}(\kappa)$. The normalized eigenfunctions of $\slashed{D}$ read
\begin{eqnarray}
&&\psi_\lambda =
\frac{e^{\ii(\vartheta+k)(\tau-\pi)}}{N_\lambda}
\bigg(\varphi^+(\kappa)+\frac{\vartheta+k-\lambda}{\kappa}\varphi^-(\kappa)\bigg)\otimes \begin{pmatrix}
1\\
\ii
\end{pmatrix}\nonumber\\
&&\qquad\qquad+\frac{e^{\ii(\vartheta+k)(\pi-\tau)}}{N_\lambda}
\bigg(\psi^+(\kappa)-\frac{\vartheta+k-\lambda}{\kappa}\varphi^-(\kappa)\bigg)\otimes \begin{pmatrix}
1\\
-\ii
\end{pmatrix},\label{ps11}
\end{eqnarray}
where $k\in\mathbb{Z}$ and $N_\lambda$ is a normalization factor,
\begin{equation}\label{ps18}
N_\lambda=\sqrt{2\pi}\Bigg(1+\frac{(\vartheta+k-\lambda)^2}{\kappa^2}\Bigg)^{1/2}.
\end{equation}
The eigenvalues corresponding to these eigenfunctions
\begin{equation}
\lambda=\pm\sqrt{\kappa^2+(\vartheta+k)^2}
\end{equation}
are symmetric with respect to the reflection and thus do not contribute to the $\eta(s,\slashed{D})$.

The zero modes $\varphi_{0,\rho}$ of $\mathcal{D}$ can be classified according to their $\Gamma^\tau$-chirality, $\mathcal{D}\varphi_{0,\rho}=0$, $\Gamma^\tau\varphi_{0,\rho}=\rho\varphi_{0,\rho}$, where $\rho=\pm 1$. The corresponding eigenmodes of $\slashed{D}$ read
\begin{equation}\label{ps17}
\psi_\lambda =
\frac{e^{\ii(\vartheta+k)(\tau-\pi)}}{\sqrt{4\pi}} \varphi_{0,\rho}\otimes
\begin{pmatrix}
1\\
\ii
\end{pmatrix}
+\rho\frac{e^{\ii(\vartheta+k)(\pi-\tau)}}{\sqrt{4\pi}} \varphi_{0,\rho}\otimes
\begin{pmatrix}
1\\
-\ii
\end{pmatrix}.
\end{equation}
For these modes, we get
\begin{equation}
\lambda = \rho\,\vartheta +k .\label{lamrho}
\end{equation}
For $\vartheta=0$, zero modes appear, but the non-zero spectrum is symmetric. Therefore,
\begin{equation}
\eta(s,\slashed{D})=0\quad \mbox{for}\quad \vartheta=0.\label{etat0}
\end{equation}

Let us call $n_\pm$ the number of zero modes of $\mathcal{D}$ with $\rho=\pm 1$ and define the $\Gamma^\tau$-index of $\mathcal{D}$ as
\begin{equation}
\mathrm{ind}\, (\mathcal{D},\Gamma^\tau)=n_+-n_-.\label{ind}
\end{equation}
Then, for $\vartheta\in (0,1)$, we have
\begin{align}\nonumber
\eta(s,\slashed{D})&=n_+\Bigg(\sum_{k=0}^{+\infty}(\vartheta+k)^{-s}-
\sum_{k=-1}^{-\infty}(-\vartheta-k)^{-s}\Bigg)+
n_-\Bigg(\sum_{k=1}^{+\infty}(-\vartheta+k)^{-s}-
\sum_{k=0}^{-\infty}(\vartheta-k)^{-s}\Bigg)\\ \label{ps14}
&=\mathrm{ind}\, (\mathcal{D},\Gamma^\tau)\big(\zeta_H(s,\vartheta)-\zeta_H(s,1-\vartheta)\big),
\end{align}
where
\begin{equation}
\zeta_H(s,\vartheta)=\sum_{k=0}^{+\infty}(\theta+k)^{-s}\label{Hurw}
\end{equation}
is the Hurwitz (generalized Riemann) $\zeta$ function. Taking into account that $\zeta_H(0,\vartheta)=\tfrac 12 -\vartheta$, we obtain
\begin{eqnarray}
&\eta(0,\slashed{D})=0\quad &\mbox{for}\quad \vartheta=0\nonumber,\\
&\eta(0,\slashed{D})=(1-2\vartheta)\,\mathrm{ind}\, (\mathcal{D},\Gamma^\tau)  \quad &\mbox{for}\quad \vartheta\in (0,1).\label{eta0varth}
\end{eqnarray}

\subsection{General statement}\label{sec:oddgen}
General setup in this section is similar to that of Section \ref{bag}. We assume that near each component of the boundary $\partial\MM_\alpha$ the bundle $V$ has a product structure in a collar neighbourhood $\partial\MM_\alpha\times [0,\tau_\alpha)$. 
\begin{equation}
\Pi_-=\tfrac 12 (1-\chi), \label{Piminodd}
\end{equation}
where $\chi$ is an idempotent, $\chi^2=1$, having identical numbers of positive and negative eigenvalues. 
We request that $\chi$ defines Clifford boundary conditions \cite{Gilkey:1983xz}, namely such that $\chi\gamma^\nn=-\gamma^\nn\chi$ and $\chi\gamma^a=\gamma^a\chi$. As we have said above, this ensures hermiticity of the Dirac operator and strong ellipticity of the boundary value problem. Moreover, we request that the boundary projector commutes with the rotations of tangential directions, $[\chi,[\gamma^a,\gamma^b]]=0$. The solution for all these conditions is essentially given by (\ref{ps2}),
\begin{equation}
\chi=\Gamma^\nn \otimes(\cos(2\pi\vartheta_\alpha)\sigma_3+\sin(2\pi\vartheta_\alpha)\sigma_1).
\end{equation}

Let us make the small variation $\vartheta_\alpha\to\vartheta_\alpha+\delta\vartheta_\alpha$. It is not obvious yet that the variation of $\eta(0,\slashed{D})$ can be expressed through the heat kernel coefficients. Thus, we extend smoothly $\delta\vartheta_\alpha$ to the bulk in such a way that it vanishes outside a small vicinity of the boundary. Next, we pass to an equivalent spectral problem by making the infinitesimal unitary transformation $\slashed{D}\to U^\dag \slashed{D}U$, $\Pi_-\to U^\dag \Pi_- U$ with $U\simeq \mathrm{id}\otimes (1-\ii \pi \sigma_2(\delta\vartheta_\alpha))$. Due to this transformation, we return to the unperturbed value of $\vartheta_\alpha$ in the boundary projector, but the bulk Dirac operator receives the variation
\begin{equation}
\delta\slashed{D}=\pi \left(\Gamma^\mu\otimes 1\right)\, \partial_\mu (\delta\vartheta_\alpha).
\label{delD}
\end{equation}
This variation satisfies the conditions formulated in Section \ref{sec:in} and thus the corresponding variation of $\eta$ function at $s=0$ is given by (\ref{vareta}). For odd dimension $n$, the heat kernel coefficient $a_{n-1}$ contains both bulk and boundary contribution. As we have mentioned above, different extensions of $\delta\vartheta_\alpha$ from $\partial\MM_\alpha$ to the bulk lead to isospectral problems and $\delta\vartheta_\alpha$ can be concentrated in an arbitrarily small vicinity of the boundary, where $\MM$ and $\slashed{D}$ have a product structure. Thus, the relevant invariants appearing in the heat kernel expansion are the ones associated with $\partial\MM_\alpha$, $\mathcal{D}_\alpha$ and the boundary angle $\theta_\alpha$. By repeating the arguments from Section \ref{sec:gen}, we conclude that the \emph{smooth} part $\eta(0,\slashed{D})$ can be written as 
\begin{equation}
\eta(0,\slashed{D})=\sum_\alpha h(\theta_\alpha,\mathcal{D}_\alpha) \label{etaodd}
\end{equation}
with some unknown function $h$ of $\vartheta_\alpha$ and geometric data associated with $\mathcal{D}_\alpha=\ii\Gamma^a\nabla_a\vert_{\partial\MM_\alpha}$. This function is determined from the smooth part of (\ref{eta0varth}),
\begin{equation}
h(\theta_\alpha,\mathcal{D})=(1-2\vartheta_\alpha) \,\mathrm{ind}\, (\mathcal{D}_\alpha,\Gamma^\nn ).\label{halpha}
\end{equation}

We remind that $\vartheta=0$ and $\vartheta=1$ correspond to identical boundary conditions. Therefore, the smoothness of (\ref{etaodd}) is modulo $2\mathbb{Z}$. As in Section \ref{bag}, relation (\ref{etaodd}) is an equation for smooth variations. It can be integrated by passing to the exponentiated $\eta$-invariant. This time the integration constant can be defined unambiguously in contrast to (\ref{Ebag}). Let us put $\vartheta_\alpha=0$. Due to the symmetry of the spectrum $\eta(0,\slashed{D})=0$ and $\eta_{\slashed{D}}=\tfrac 12 \mathcal{N}_0$, where $\mathcal{N}_0$ is the number of zero modes of $\slashed{D}$, when all $\vartheta_\alpha=0$. Then, for general $\vartheta_\alpha$ we obtain 
\begin{equation}
\mathcal{E}(\slashed{D})=\exp\left(-\pi \ii \mathcal{N}_0 +2\pi \ii \sum_\alpha \vartheta_\alpha   \,\mathrm{ind}\, (\mathcal{D}_\alpha,\Gamma^\nn )\right) .\label{Eodd}
\end{equation}

Note, that the coefficient in front of $\mathrm{ind}\, (\mathcal{D}_\alpha,\Gamma^\nn )$ in (\ref{halpha}) is not integer and not even a half-integer. Expression (\ref{halpha}) reminds the $\eta$ function computed in \cite{MateosGuilarte:2019eem} for domain walls in 3D, though the angle $\vartheta$ plays there a different role.

\section{Conclusions}\label{sec:con}
Let us summarize the main features of our method. The first step is to analyse the heat kernel expansion and to define the generic form of the smooth part of $\eta(0,\slashed{D})$ in terms of a universal function of the boundary Dirac operator $\mathcal{D}$. Then we use exactly solvable examples in the spirit of \cite{Gilkey:1983a} to determine these universal functions. The smoothness assumption can be removed by passing to the exponentiated $\eta$-invariant. In even dimensions we obtained expressions for $\mathcal{E}(\slashed{D})$ in terms of the $\eta$-invariants of boundary operators, see (\ref{Ebag}), while in odd  dimensions $\mathcal{E}(\slashed{D})$ is expressed through the index of boundary Dirac operators, see (\ref{Eodd}). 

Let us address the question which generalizations are possible and which assumptions can be lifted. The key assumption, which is quite common in this field, is the product structure near the boundary. As the computations in low dimensions show, see Section \ref{sec:two} and \cite{Kurkov:2017cdz}, this assumption on the gauge field can probably be weakened. However, the metric needs to have a product form since non-vanishing extrinsic curvature of the boundary is a serious obstacle to the interpretation of the $\eta$ function in terms of boundary operators \cite{Kurkov:2018pjw}. Adding more fields, like scalar or axial vector fields, may also be problematic.

On a practical side, the anomaly inflow mechanism allows us to formulate anomaly-free theories of boundary fermions \cite{Witten:2019bou}. We see somewhat similar effects for our boundary conditions as well. For some combinations of fermions on $\MM$ and $\partial\MM$ there are cancellations between ambiguous phases generated by the anomalies in path integrals. However, an important ingredient is still missing. One has to interpret the boundary theories in terms of edge states of $\slashed{D}$. For example, a single Dirac fermion with bag boundary conditions may have one or no edge states depending on the sign of the mass, while the coefficient relating the $\eta$ functions in the massless case is $1/2$, see (\ref{fD}). The fact that the massless case appears exactly halfway between positive and negatives masses gives us a certain degree of optimism, but it also indicates that massive fermions have to be included in our formalism.

The inclusion of masses is also necessary to compare the heat kernel methods to that of Ref.\ \cite{Witten:2019bou}. In principle, both large and small mass expansions of the $\eta$ function can be analysed with the help of the heat kernel \cite{Deser:1997gp} though require considerable technical efforts. Note that direct calculations of the boundary Hall conductivity within Pauli--Villars regularization scheme \cite{Kurkov:2020jet} are consistent with the heat kernel calculations of the $\eta$ function.

\acknowledgments
We are grateful to Edward Witten and Kazuya Yonekura for very useful explanations of their boundary conditions and to Maxim Kurkov for collaboration on related topics. The work of A.V.I. is supported by the Ministry of Science and Higher Education of the Russian Federation, agreement  075-15-2022-289. The work of D.V.V. was supported in parts by the S\~ao Paulo Research Foundation (FAPESP), grant 2021/10128-0, and by the National Council for Scientific and Technological Development (CNPq), grant 305594/2019-2.

\bibliographystyle{JHEP}
\bibliography{graphene,parity}

\end{document}